\documentclass [prl, twocolumn, aps, superscriptaddress
] {revtex4}
\usepackage{graphicx}
\usepackage{amsmath}
\usepackage{amssymb}
\usepackage{color}

\newcommand{\be}{\begin{equation}}
\newcommand{\ee}{\end{equation} }
\newcommand{\ba}{\begin{eqnarray} }
\newcommand{\ea}{\end{eqnarray} }
\newcommand{\n}{\nonumber \\ }

\begin{document}

\title{Devil's staircases and supersolids in a one-dimensional dipolar Bose gas}

\author{F. J. Burnell}
\affiliation{Department of Physics, Princeton University, Princeton,
New Jersey 08544, USA} %

\author{Meera M. Parish}
\affiliation{Department of Physics, Princeton University, Princeton,
New Jersey 08544, USA} %
\affiliation{Princeton Center for Theoretical Science, Princeton
University, Princeton, New Jersey 08544, USA}

\author{N. R. Cooper}
\affiliation{Cavendish Laboratory, JJ Thomson Avenue, Cambridge, CB3
0HE, United Kingdom}

\author{S. L. Sondhi}
\affiliation{Department of Physics, Princeton University, Princeton,
New Jersey 08544, USA} 

\date{\today}

\begin{abstract}
We consider a single-component gas of dipolar bosons 
confined in a one-dimensional optical lattice, where the dipoles are
aligned such that the long-ranged dipolar interactions are maximally
repulsive. In the limit of zero inter-site hopping and sufficiently
large on-site interaction, the phase diagram is a complete devil's
staircase for filling fractions between 0 and 1, wherein every
commensurate state at a rational filling is stable over a finite
interval in chemical potential. We perturb away from this limit in
two experimentally motivated directions involving the addition of
hopping and a reduction of the onsite interaction. The addition of
hopping alone yields a phase diagram, which we compute in
perturbation theory in the hopping, where the commensurate Mott
phases now compete with the superfluid. Further softening of the
onsite interaction yields alternative commensurate states with
double occupancies which can form a staircase of their own, as well
as one-dimensional ``supersolids'' which simultaneously exhibit
discrete broken symmetries and superfluidity.
\end{abstract}

\maketitle

The unprecedented control over experimental parameters in trapped
ultracold atomic gases has substantially widened the range of
possible exotic phases of matter that can be explored. Optical
lattices can be used to simulate simple lattice models and/or vary
the system dimensionality, while the interatomic interactions can
be varied via a magnetically-tunable Feshbach
resonance~\cite{bloch2008}. Thus far, the focus has largely been on
\textit{contact} interactions, since these generally provide a
good description of atom-atom scattering in the low energy limit.
However, with the realization of atomic gases with strong magnetic
dipole moments~\cite{lahaye2007} and the prospect of working with
molecules that exhibit electric dipole
moments~\cite{sage2005,Ni2008}, there is now substantial interest in
examining the physics of long-ranged interactions in these
systems ~\footnote{For a review of the current understanding of dipolar interactions, and the parameter r{\'e}gime currently accessible to experiments, see ~\cite{DipoleReview, DipoleReview2}}.

One important feature of long ranged interactions is that they can
produce exceptionally intricate ground-state structures in the
classical, or strong coupling, limit. A particularly elegant example
of this is the case of classical particles in a one-dimensional (1D)
lattice interacting via an infinite-ranged convex potential studied
by Pokrovsky and Uimin, and Hubbard
(PUH)~\cite{Hubbard,PokrovskyUimin,Pokrovsky}. Here, it can be shown
that the ground state filling fraction as a function of chemical
potential $\mu$ is a complete devil's staircase~\cite{Bak}, in which
every rational filling fraction between 0 and 1 is stable over a
finite interval in $\mu$, and the total measure of all such
intervals exhausts the full range of $\mu$.

In this work, we show that the devil's staircase of PUH has dramatic
consequences for the physics of quasi 1D cold atomic gases. Building
on the existing understanding of this classical limit, we consider
two perturbations of the devil's staircase that arise naturally in
the experimental setting of cold atomic gases. The first of these is
the introduction of a quantum kinetic energy which now renders the
problem sensitive to particle statistics---which we take to be
bosonic, given that this case can currently be realized with either
atoms or molecules. The second perturbation involves tuning the
onsite interaction independently of the rest of the interaction.
This allows for a controlled departure from convexity, and hence
from the PUH states considered previously. The first perturbation
has the well understood effect of initiating a competition between
the crystalline, Mott phase that exists at zero hopping and the
superfluid (Luttinger liquid in $d=1$) that must exist at all
fillings at sufficiently large hopping. By means of strong coupling
perturbation theory similar to that previously studied in the
Bose-Hubbard model~\cite{FreericksMonien}, and in extended
Bose-Hubbard models with nearest-neighbor
interactions~\cite{WhiteMonien, ScalletarBH1}, we derive a phase
diagram that exhibits this evolution and which we supplement by
standard wisdom from the Luttinger liquid description of the
transitions. The second perturbation introduces doubly occupied
sites in the classical limit. While describing the resulting phase
diagram in complete and rigorous detail is beyond the approaches we
take in this paper, we give an account of the ``staircase''
structure of the initial instability and of regions of the phase
diagram where the classical limit states exhibit superlattices of
added charge built on underlying PUH states. At least some of these
regions exhibit devil's staircases of their own. Finally, upon
introducing hopping we are led to an infinite set of
``supersolids''---which in this context are phases that are both
Luttinger liquids and break discrete translational symmetries.

\noindent {\bf Model:} The Hamiltonian we will study is:
 \ba \label{HHup}
 H &= &V_0 \sum_{i< j}\frac{1}{r_{ij}^3} n_i n_j + \frac{U}{2} \sum_{i} n_i (n_i-1) \n
 &&- \mu \sum_i n_i -t \sum_{i} c^\dag_i c_{i+1} +h.c.
 \ea
This model describes bosons in a deep 1D optical lattice with
hopping amplitude $t$, on-site interaction energy $U$, and an
infinite-ranged dipole-dipole interaction $V_0/r^3$.  In cold-atom
systems the hopping potential $t$ is controlled by the depth of the
optical lattice, and hence can be tuned over a wide range of values.
In the cases where the bosons are atoms, e.g.\
$^{52}$Cr~\cite{lahaye2007}, the on-site interaction $U$ may also be
easily tuned using Feshbach resonances. We set the dipolar
interaction to be maximally repulsive by aligning the dipoles
perpendicular to the 1D chain.

If $U$ is sufficiently large, the potential is everywhere
convex~\footnote{A potential is convex if it obeys $ V(x) \leq
\lambda V(x-1+\lambda) + (1-\lambda) V( x+ \lambda )$
 for $0 \leq \lambda \leq 1$.}
and the $t=0$ (classical) ground states of (\ref{HHup}) are those of
a PUH Hamiltonian. For every rational filling fraction $\nu= p/q$,
the ground state is periodic with period $q$~\cite{Hubbard}.  Each
such ground state is unique up to global lattice
translations~\cite{Burkov}. We denote these states {\it commensurate
ground states} (CGS). Adding or removing a single particle from a
CGS in the infinite volume limit produces a {\it
 q-soliton state} (qSS) containing $q$ fractionally charged solitons
of charge $1/q$. Each soliton is a distortion of the periodic
ground state which alters the distance between one pair of adjacent particles by 1.
(For every commensurate state, there is a unique distortion of this type which minimizes
the potential energy). The range of $\mu$ over which each CGS is
stable is given by
 \be
 \sum_{n=1}^\infty \frac{n q}{(n q+1)^3} +\frac{n q}{(n q-1)^3} - \frac{2 n q}{(n q)^3}
 \ee
and falls off sharply as a function of $q$.  At $t=0$ these
intervals cover the entire range of $\mu$ pertinent to fillings less
than unity, giving the devil's staircase structure.

We first direct our attention to the $t \ne 0$ phase portrait but
still at large $U$, where the potential is everywhere convex. The
qualitative behavior of our system in this regime is reminiscent of
the Bose-Hubbard model, with commensurate Mott lobes ceding to
superfluid states as $t$ increases.  A state with large $q$ may be
treated as a state with smaller $q$ at a nearby filling in which a
crystal of dilute solitons has formed. Hopping tends to liquify
dilute crystals of solitons: at large separation the inter-soliton
repulsion is smaller than the kinetic energy gained from
delocalization.  The delocalized solitons destroy long-ranged
spatial order, creating a Luttinger liquid with full translational
symmetry. Hence, as $t$ increases, the system undergoes a transition
from the Mott insulating CGS to a Luttinger liquid state, with
larger $q$ states liquifying at smaller $t$.

\noindent {\bf Strong Coupling Expansion: } To find the position of
the phase boundary, we generalize the method of
Ref.~\cite{FreericksMonien} and compare the energies of the CGS and
its adjacent qSS to third order in $t$  using standard
time-independent perturbation theory. This approach assumes that the
phase transition is continuous, so that for a given $t>0$, values of
$\mu$ for which $E_{qSS}(\mu) = E_{CGS}(\mu)$ constitute the phase
boundary. This assumption is well-founded, since the soliton
repulsion ensures that the energy cost of creating multiple solitons
is larger that of a single soliton, thus favoring a second-order
transition.

In a finite system one must account for the repulsion between
solitons; here we drop these terms and consider the infinite
volume limit. In the limit that the solitons are sufficiently well
separated that we may neglect their interactions, the qSS is
highly degenerate and can be expressed in terms of a band of solitonic
momentum eigenstates. Here we consider only the
bottom of the band, which lies at zero momentum.

The first and third order corrections to the CGS energy are zero.
At $\nu = p/q$, the second-order correction is given by
 \ba
  E^{(2)} &=& - 2 \frac{N t^2}{p} \sum_{i=1}^p \frac{1}{\Delta E_i}
 \ea
where $\Delta E_i = E_i^{(0)}-E_0 ^{(0)}$ is the difference in
potential energies between the ground and the excited state formed
by hopping from the $i^{th}$ occupied site in the ground state
configuration.  As the ground state is periodic, it suffices to
calculate these energies for the $p$ distinct particles in the
repeated pattern.

For the qSS at $\nu = p/q$, we consider the infinite volume
limit, in which the total energy correction is simply $q$ times
the energy correction for a state with a single soliton.  The
soliton hops by $q$ sites when a single boson on one of its edges
is hopped by one site. The resulting energy corrections are
 \ba
E^{(1)} &=& -2 qt  \cos (kq a )  \n
 E^{(2)} &=& 2 q \cos(2 k q a) \frac{t^2}{\Delta E_{r_1, -1}}
 - q \sum_{i=1}^{N/q} \sum_{\alpha=\pm 1} \frac{t^2}{\Delta
  E_{r_i, \alpha}}\n
 E^{(3)} &=&- 2 q t^3 \cos(k q a)  \left [  \frac{\cos (2 k q
a)}{\Delta E_{r_{1, -1}} \Delta E_{r_{1, -1}+q}} - \frac{\cos (2 k
q a)}{(\Delta E_{r_{1, -1}})^2}  \right ]\n
 & & -2q t^3 \cos(kqa)
\sum_{i=1}^{N/2} \left [\frac{1}{(\Delta E_{r_i})^2} -
\frac{1}{\Delta E_{r_i} \Delta E_{r_{i+1}}} \right ]
 \ea
Here $k$ is the soliton momentum, $a$ is the lattice constant, and
$q$ is the denominator of the commensurate filling fraction, which
appears here because one hopping displaces the soliton by $q$
lattice spacings.  The subscripts on $E_{r_i, \pm1}$ indicate the
distance between particle $i$ and the soliton, and the direction
of the hopping relative to the soliton. The special distance
$r_{1} $ describes an excited state in which an anti-soliton is
sandwiched between 2 solitons, or vice versa.  These states
contribute extra terms to the energy corrections of the qSS
because of the ambiguity as to which soliton is associated with
the ground state qSS.

\begin{figure}[htp]
\begin{center}
\includegraphics[width=3.4in, height=1.9in]{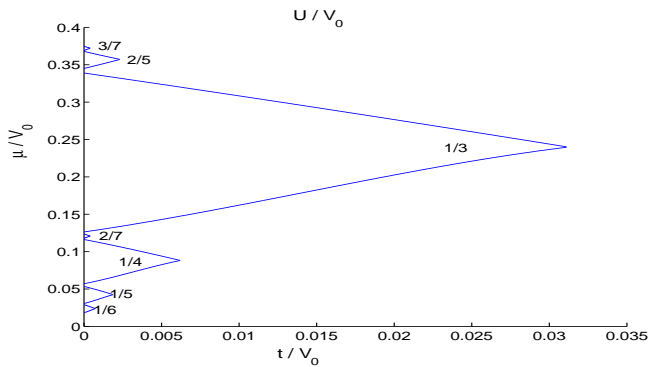}
\end{center}
\caption{\label{Hoppings} Perturbative calculation of the Mott to SF
phase boundary in the $(t, \mu)$ plane, shown here for $U=20$.  Here
the strength of all couplings is measured relative to that of the
dipolar interaction strength $V_0$.  Each lobe encloses a Mott
insulating region in which the filling is fixed; the region outside
the lobes is a superfluid of solitons.  The $1/3, 1/5, 1/6, 2/7,
2/7$ and $3/7$ -filled lobes are shown here.  Every commensurate
state of the complete Devil's staircase has a Mott lobe, but the
range of hoppings over which a state exists falls off sharply with
its denominator. }
\end{figure}

Fig.~\ref{Hoppings} shows the results of the perturbative
calculation for selected filling fractions.  The $t=0$ axis
corresponds to the classical limit, in which the CGS states
comprise a complete devil's staircase: every value of $\mu$
corresponds to a rationally filled ground state, except for a set
of measure 0.   The figure shows the resulting Mott lobes: inside
each lobe the CGS is stable and the system is in a Mott insulating
state.  The Mott gap vanishes on the boundary of the lobe;
outside of this region solitons proliferate and the system is in a
Luttinger liquid phase.

For any $t>0$, only a finite number of insulating states exist; the
rest are liquid states with a superfluid of condensed solitons. The
function $\nu (\mu)$ is no longer a devil's staircase, but rather a
piecewise smooth function, with plateaux of constant density
separated by liquid phases whose density varies continuously with
$\mu$.   The size of the commensurate region decreases sharply with
$q$: states of higher $q$ have both smaller ranges of stability in
the classical limit, and larger energy corrections relative to the
CGS, so that the volume of the corresponding Mott lobe scales
approximately as $1/q^5$\footnote{ This scaling uses a first-order
in $t$ approximation to the boundaries of the Mott lobe.}. The total
volume occupied by liquid states can be estimated from the
first-order approximation to the Mott lobe boundaries; we find that
for small but fixed $t$ the volume of the liquid region scales as
approximately $t^{2/5}$.  At sufficiently large $t$ we expect all
insulating states to be unstable, and the particle density to vary
smoothly with $\mu$.

Thus far, our results only apply to a homogeneous system, but
we can estimate the effects of a harmonic trapping potential
(present in current cold-atom experiments) using the local density
approximation -- this assumes that the trapping potential is slowly
varying enough that it can be simply incorporated into the chemical
potential, resulting in a spatially-dependent $\mu$. Trajectories
along the 1D chain then correspond to cuts in Fig.~\ref{Hoppings} at
fixed $t/V_0$. Thus, at $t=0$ we expect different commensurate
fillings at different points along the trapped chain, with the most
stable states (fillings $1/2$ and $1/3$) occupying the largest
regions within the trap. Note, however, that commensurate states
with a period $q$ greater than the lengthscale over which the
trapping potential varies do not exist in the trapped system since
they violate the local density approximation. Moreover, once $t$ is
nonzero but small, we have small islands of Mott states at various
fillings $p/q$, separated by regions of superfluid which interpolate
continuously between the two commensurate densities.

The qualitative nature of the Mott transition can be deduced from
existing knowledge of 1D commensurate-incommensurate phase
transitions, which we summarize here. Bosonization can be used to
treat the kinetic term and dipolar interactions exactly; the lattice
potential must be treated perturbatively. In bosonized form, our
system is described by a Hamiltonian that is a sum of
a Luttinger piece, accounting for the kinetic term and dipolar
interactions~\footnote{Though the potential is infinite ranged, it
falls off quickly enough that the qualitative description is
identical to that for short-ranged interactions~\cite{Giamarchi}.},
and a sine-Gordon term which emulates the lattice in the continuum
limit.

The resulting Hamiltonian is well studied in both the context of the
Mott transition \cite{Giamarchi} and the Frenkel-Kontorowa model of
surface interfaces \cite{Schulz}.  The upper and lower Mott lobes
join in a cusp which is not accurately described by the perturbation
theory-- since `small' hopping implies that the ratio $t/ \delta
E_i$ of the hopping relative to the energy gap to the solitonic
states must be small, the range of $t$ over which the perturbative
treatment is valid decreases with $q$ and never encompasses this
point.  The bosonized treatment reveals that crossing the edge of
the Mott-Hubbard lobe induces one of two different types of phase
transitions. At the cusp joining the upper and lower Mott lobes, a
constant-density phase transition of the Kosterlitz-Thouless type
occurs. Everywhere else the transition is well described by a simple
two-band model with quasi-particles that are gapped in the
commensurate phase, and with a density increasing as $\sqrt{ \mu -
\mu_c}$ near the transition on the liquid side.

\begin{figure}[htp]
\begin{center}
 \includegraphics[width=3.4in, height=1.9in]{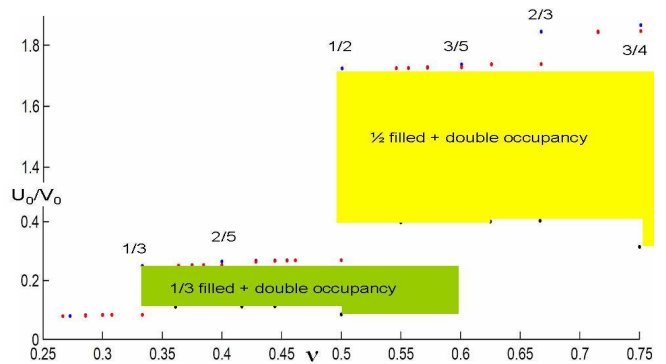}
\end{center}
 \caption{\label{U0s}(color online) Numerically calculated values of $U_c^{(CGS)}$ (red) and
$U_{c}^{(qSS)}$ (blue), for a selection of filling fractions
$\nu$.  Quoted values of $U$ are measured relative to $V_0$.  A segment of the $y$-axis between $U_c^{(CGS)}(1/2)$ and
$U_c^{(qSS)}(1/2)$ has been removed for better resolution of the
rest of the phase diagram.  The green and yellow boxes, bordered
by black dots, indicate the approximate regions where the ground
states consist of double occupancies in the $1/3$ and $1/2$-filled
states respectively. Between the shaded regions the ground states
consist of double occupancies on other, higher-denominator states.
 }
 \end{figure}

\noindent
{\bf Away from the convex limit:}
The PUH CGS are the classical ground states so long as the
potential is everywhere convex.  Since the on-site potential $U$
is tunable experimentally, it is interesting to ask what
happens to these states as $U$ is lowered away from convexity and
double occupancies begin to form. Of course, as $U$ is lowered still
further, triple and higher occupancies will also form but we will not
push our analysis that far---the doubly occupied regime has enough
challenges of its own. Our results on this are summarized in Fig.~(\ref{U0s})
and we now describe the analysis behind these.

Let $U_{c}^{(CGS)}(\nu)$ be the thresholds at which the PUH ground
states give way to ones with at least one double occupancy---these
are marked as the red points in Fig.~(\ref{U0s}). Observe that these
thresholds increase monotonically with $\nu$, $U_c^{(CGS)} ( \nu') >
U_c^{(CGS)} (\nu)$ for $\nu' > \nu$. This behavior can be
qualitatively understood by considering the implications of
convexity at a given filling. Sufficient conditions for
convexity~\cite{Hubbard} are that, for all $x$,
 \ba \label{Convex}
\frac{1}{2} (V(0)+ V(2 x) )  \geq V(x)\  {\rm and} \
 U \geq\frac{15}{8} V_0 \left (\frac{1}{x^3} \right )
 \ea
For convexity to hold everywhere, (\ref{Convex}) must hold for
$x=1$; below this threshold double occupancies may occur. However,
when perturbing about a given convex solution at fixed $\mu$,
solutions will be stable approximately until $U$ violates
(\ref{Convex}) {\it for $x =r_{m}$}, the minimal inter-particle
distance. This implies that states with lower
filling fractions are more stable against double occupancies, as
the potential gain in lattice energy from doubly occupying a site
is smaller. This is in contrast to the stability of the
commensurate states as $t$ increases, where the denominator of the
filling fraction determines stability.

While our primary interest is in ground state transitions, much insight
is gained by also computing a second threshold at a given filling. Past
this threshold,  $U=U_c^{(qSS)}$, a particle added to the state goes in
as a double occupancy instead of fractionalizing into $q$ solitons. From
our above considerations, at fixed filling, the CGS is always more stable
against forming double occupancies than the qSS; a simple estimate suggests
that the gap separating these
instabilities is approximately the difference in energies for adding and
removing a particle to the state-- which is equal to the
interval in $\mu$ over which the state is stable at infinite $U$.

Now consider a filling $\nu' > \nu$. As $\nu'$ can be constructed
by adding charge to $\nu$, we conclude that $U_c^{(CGS)}(\nu') >
U_c^{(qSS)}(\nu)$; the latter corresponds to the threshold at
which adding an extra filling $\nu' - \nu$ in the form of the
solitons of $\nu$ loses out to adding it in the form of double
occupancies on top of the PUH state at $\nu$. As filling factors
only slightly greater than $\nu$ involve a dilute addition of
charges, we further conclude that $\lim_{\nu' \rightarrow \nu^+}
U_c^{(CGS)}(\nu') = U_c^{(qSS)}(\nu)$. A somewhat more involved
argument shows that $\lim_{\nu' \rightarrow \nu^-}
U_c^{(CGS)}(\nu') = U_c^{(CGS)}(\nu)~$\footnote{The argument is as
follows: we have $\lim_{\nu' \rightarrow \nu^-} U_c^{(qSS)}(\nu')
= U_c^{(CGS)}(\nu)$, as the state at filling $\nu-\delta$ is the
state at filling $\nu$ with a density $~\delta$ of hole-like
solitons; as this density decreases a single particle added as a
double occupancy will induce a charge configuration increasingly
similar to that of forming a double occupancy in the half-filled
state (which can be thought of as forming q hole-like solitons by
removing a single particle, then re-inserting this particle as a
double occupancy and letting the charge settle into its optimal
distribution.)  Further, the gap between $U_c^{(qSS)}(\nu')$ and
$U_C^{(CGS)} (\nu')$ vanishes as the denominator of the state
$\nu'$ goes to infinity, implying continuity of $U_C^{(CGS)}$ from
the left.  }

Our considerations so far imply intricate behavior for the location of
the initial ground state instability, namely that $U_c^{(CGS)}(\nu)$ is
a monotone increasing function of $\nu$ on the set of rationals with a
discontinuity at each rational value of $\nu$:
$$
\lim_{\nu' \rightarrow \nu^+} U_c^{(CGS)}(\nu') > U_c^{(CGS)}(\nu)
=\lim_{\nu' \rightarrow \nu^-} U_c^{(CGS)}(\nu').
$$
We can understand the coarse features of this instability  by considering
first the most stable states (of denominator $q \leq q_0$, for some $q_0$ small
enough to allow the exact thresholds to be computed), and deducing
the expected behavior at fillings close to these. Fig.~\ref{U0s}
plots $U_c^{(qSS)}$ (blue) and $U_c^{(CGS)}$ (red) for states with
$q \leq 15$ in the vicinity of half-filling. The values of $U_c^{(qSS)}$
shown there are obtained by numerical minimization in the sector with one
added charge at the specified filling. The values of $U_c^{(CGS)}$ are
obtained by numerical minimization over configurations {\it at} the specified
filling that contain exactly
one double occupancy. This is the correct answer for all $\nu =p/q$
for which $p$ and $q$ are not both odd. In such cases the double occupancy
and its surrounding charge rearrangements give rise to even moments starting
with a quadrupole. Consequently, double occupancies repel at all distances
and enter via a continuous transition at the computed threshold. However,
when $p$ and $q$ {\it are} both odd, the double occupancy has a dipole moment
and the transition goes first order which causes the true $U_c^{(CGS)}$ to lie
above our numerically determined value. This gap is not very large though, and
vanishes at large $q$ at least as $1/q^3$ and we ignore it here.

At any given filling, tracking the evolution of the ground state
with decreasing $U$ after double occupancies have been introduced is
a problem of considerable complexity.  Here we use the ideas
developed thus far to identify a family of regions in the $(\nu,U)$
plane where simpler descriptions emerge---these are indicated, in
two simple cases, by the shaded regions on the figure. The basic
idea is that once the $\nu = p/q$ qSS becomes unstable to double
occupancy, any particles added to the $\nu= p/q$ state will be added
as double occupancies, since these repel less strongly than
solitons.  Hence at first sight we expect, in the region
$U_c^{(qSS)}(p/q) > U > U_c^{(CGS)}(p/q)$,  states of filling $\nu >
p/q$ to consist of double occupancies in the $\nu = p/q$ state.  At
rational fillings the double occupancies will arrange themselves in
a crystal thus generating a commensuration distinct from that of the
underlying $p/q$ state---we will refer to these as
doubly-commensurate states. However, for any given $\nu$ this
argument becomes less reliable as the density of added charge
increases, since the background parent state itself becomes less
stable to forming extra double occupancies.  This can lead to
transitions in which the structure of the CGS collapses to a crystal
of double occupancies over a background of significantly smaller
filling.  We have carried out a simple analysis of the location of
this instability for the $1/2$ and $1/3$ plus double occupancy
regions in Fig.~(\ref{U0s}) at selected fillings and these are
marked by the black dots in the figure
\footnote{The result is not monotonic in filling, as once at least
half the sites are doubly occupied, the dipolar interaction of an
extra double occupancy with the existing DO's becomes comparable to
the energetic gain of creating hole-like solitons in the CGS.}.
While this indicates that there are sizeable regions which can be
described as simple descendants of the $1/2$ and $1/3$ states, we
are not at present able to estimate the sizes of analogous regions
for higher denominator fractions. Of course, to have a full solution
of this would be equivalent to tracking the evolution of each $\nu$
as $U$ is decreased from $U_c^{(CGS)}$.

As the barrier to triple occupancies is $3 U$, instabilities
towards triple occupancy at a given filling will set in at
approximately one third the value of $U$ for instabilities to
double occupancy. For $U$ close to $U_c^{qSS}(1/2)$, triple
occupancies will not be favored at any filling fraction.

\noindent {\bf A new staircase:} The above discussion has led us to
the doubly-commensurate states in the $(\nu,U)$ phase diagram: we
remind the reader that such states are constructed by periodically
doubly occupying some fraction of the sites in a CGS. We now observe
that in at least some cases these doubly commensurate states can
form a devil's staircase of their own.

First consider states constructed from double occupancies in the
$1/2$-filled state, which exist in the region shaded in yellow in
Fig.~(\ref{U0s}). The energetics of such states can be divided into
a) the constant interaction of the parent $1/2$-filled PUH
configuration with itself, b) the constant interaction of the added
charges, irrespective of their location, with the parent $1/2$
filled configuration and c) the interaction of the added charges
with themselves. This last part involves an interaction between the
added charges which is convex again and thus leads to PUH
configurations sitting on a lattice with a doubled lattice constant.
The energy cost of adding a single double occupancy is $U+V_d$,
Hence the doubly-occupied sites comprise a Devil's staircase with
$\mu \rightarrow \mu +U + V_d$, and the widths of all intervals
decreased by a factor of 8. Here $V_d
=\frac{1}{8}\sum_{n=1}^{\infty} \frac{1}{n^3}$ is the interaction
energy of each double occupancy with the underlying $1/2$-filled
state.  At fixed $U$, this staircase is complete over the range of
fillings for which increasing the particle density infinitesimally
does not induce `excess' double occupancies to form in the
half-filled background lattice.  In the case of the $1/2$-filled
state, for $U$ sufficiently close to the upper cutoff this gives a
complete staircase on $1/2 \leq \nu \leq 1$.  Similar structures
exist for all $1/q$-filled states in the appropriate range of $U$.
As mentioned before, we do not, at present, understand the situation
for doubly commensurate descendants of general rational fillings.

\noindent
{\bf ``Supersolids'':}
Thus far our considerations away from the convex limit have been purely
classical. But we can equally consider states obtained from these modified
classical states upon the introduction of hopping. Specifically, let
us consider the fate of the doubly commensurate descendants of the PUH
$1/q$ states considered above.

In a manner entirely analogous to the problem with which we began
this paper, the superlattice of added charges can melt via the
motion of {\it its} solitons as $t$ is increased resulting in a
phase transition between the doubly commensurate state and a
``super-solid'' like phase in which the background $1/q$-filled CGS
coexists with a Luttinger liquid. This is, in a sense the $d=1$
version of the supersolid in higher dimensions, but it is worth
noting that the $d=1$ version in our problem exhibits a more
divergent CDW susceptibility than superfluid susceptibility as $T
\rightarrow 0$.

\begin{figure}[htp]
\begin{center}
 \includegraphics[width=3.4in, height=1.9in]{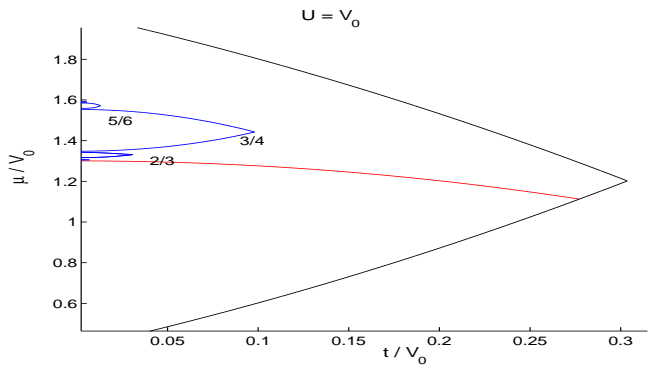}
 \end{center}
\caption{\label{SuperSolid}(color online) Phase portrait in the
vicinity of doubly commensurate states about $1/2$ filling at $U
=V_0$.  Again, $\mu, t,$ and $U$ in the figure are measured relative to $V_0$.
The black lines indicate the boundaries of the Mott lobe at
$U = 20$.  The red curve shows the chemical potential at which
it becomes energetically favorable to add particles to the
half-filled state as double occupancies, for $U=V_0$.  The blue
curves show some Mott lobes of the doubly-occupied staircase
region; the region between these and the black lines (which
delineate the region of stability of the $1/2$-filled state at
infinite $U$) is a super-solid state. }
 \end{figure}

To get a more quantitative account of these new phases, we may
repeat the strong coupling treatment above. Fig.~(\ref{SuperSolid})
shows the phase portrait at intermediate values of $U$ near
$\nu=1/2$. The black line traces the infinite-$U$ Mott lobe, over
which the background $1/2$-filled state is stable against forming
solitons. The red line shows the threshold at which it is
energetically favorable to add a single double occupancy to the
$1/2$ filled state.  The blue curves show the positions of the Mott
lobes for the doubly commensurate states. The presence of double
occupancies stabilizes the $1/2$-filled state against proliferation
of solitons, so that the background remains commensurate at least
within the infinite-$U$ $1/2$-filled Mott lobe, shown in black.
This $1/2$-filled super-solid phase has also been shown to exist in
an extended Bose-Hubbard model with second-neighbor
repulsion~\cite{ScalletarSuperSolid}; these numerical results are
consistent with the phase portrait shown here for small $t/V$.

\noindent
{\bf Concluding Remarks:} The long range of the dipolar interaction
does produce, as promised, intricate phase diagrams for the one
dimensional dipolar bosonic gas. Particularly striking are the singular
staircase functions that showed up in our analysis in three different
settings: in the $\nu(\mu)$ curve for the exactly dipolar classical
problem, in the function $U^{(CGS)}_{0c}$ which marks the instability
of the PUH states when the onsite $U$ is tuned down and in the
$\nu(\mu)$ curves in selected regions of the $(\nu, U)$ plane. The
other main set of results pertain to the presence of a large, indeed,
infinite number of transitions between Mott crystals and Luttinger
liquids or supersolids. The challenge of observing some of this physics
in cold atomic gases is not trivial---the major obstacles are getting
a reasonable simulacrum of a one dimensional gas of infinite extent. On
the positive side, the control parameters we study here are eminently
tunable and when combined with the physics we describe above, we hope,
will motivate our experimental colleagues to invest some time in performing
the needed miracles.

\noindent
{\bf Acknowledgements} The authors wish to thank David Huse
and Steven Kivelson for numerous helpful discussions and the latter,
especially, for introducing us to the work of Bak and Bruinsma.  NRC gratefully acknowledges the 
support of ICAM, and SLS acknowledges the support of NSF Grant No. DMR 0213706.


\end{document}